\documentclass[twocolumn,twoside,slac_two]{revtex4}
\usepackage{graphicx}
\usepackage{fancyhdr}
\usepackage{amsmath}
\begin{document}

\title{Determining the location of the GeV emission in powerful blazars}

\author{Amanda Dotson}
\affiliation{Department of Physics, Joint Center for Astrophysics,
University of Maryland Baltimore County, 1000 Hilltop Circle,
Baltimore, MD 21250, USA}

\author{Markos Georganopoulos}
\affiliation{Department of Physics, Joint Center for Astrophysics, 
University of Maryland Baltimore County, 1000 Hilltop Circle, Baltimore, MD 21250, USA
\\
NASA Goddard Space Flight Center, Code 660, Greenbelt, MD 20771, USA}

\author{Demosthenes Kazanas}
\affiliation{NASA Goddard Space
Flight Center, Code 660, Greenbelt, MD 20771, USA}

\author{Eric Perlman}
\affiliation{Department of Physics and Space Sciences,
Florida Institute of Technology, 150 West University Boulevard,
Melbourne, FL  32901, USA}

\begin{abstract}
 	An issue currently under debate in the literature is how far from the black hole is the {\sl Fermi}-observed GeV emission of powerful blazars emitted. Here we present a diagnostic tool for testing whether the GeV emission site is located within the sub-pc broad emission line (BLR) region or further out in the pc scale molecular torus (MT) environment.  Within the BLR the scattering takes place at the onset of the Klein-Nishina regime, causing the electron cooling time to become almost energy independent and as a result, the variation of high-energy emission is expected to be achromatic.  Contrarily, if the emission site is located outside the BLR, the expected GeV variability is energy-dependent and with amplitude increasing with energy.  We demonstrate this using time-dependent numerical simulations of blazar variability and discuss the applicability of our method. 
\end{abstract}

\maketitle

\thispagestyle{fancy}

\section{Introduction}
\label{sec:intro}
	Blazars are by far the most common objects detected in the gamma-ray sky \citep{abdo112fgl}. \textit{Fermi} has detected blazar variability as short as a few hours \citep[e.g.][]{abdo10b}. During these flares, the GeV luminosity has been known to increase by a factor of up to several compared to its pre-flare luminosity.  Because blazars cannot be resolved at these energies (or at any other energy, with the possible exception of VLBA observations), it is impossible to determine the location of these flares by direct detection.  To address this issue, we propose a diagnostic test that utilizes \textit{Fermi} variability data of short flares to determine the location of the GeV emission in high-power blazars, namely flat spectrum radio quasars (FSRQ).  
	
\section{Sources of Seed Photons}	
	
	 Relativistic effects determine which photon field is dominant at varying distances from the central black hole, and as a result the location of the GeV flaring site determines the dominant source of seed photons.  The co-moving (jet frame) energy density of a radiation field $U'$ scales as differing factors of $\Gamma$ depending on the direction from which the photons enter the emitting region \citep{dermer94,geo01}. If the photon field is isotropic 
	 \begin{equation}
	 \label{eq:uiso}
	 U' \approx \Gamma^2 U = \frac{ \Gamma^2 L}{4 \pi R^2 c} 
	 \end{equation}

For photons entering the emitting region from behind the relativistically moving blob 

	 \begin{equation}
	 \label{eq:ubhd}
	 U' \approx \Gamma^{-2} U = \frac{L}{\Gamma^{-2} 4 \pi R^2 c} 
	 \end{equation}
	 
In this work, all primed quantities refer to the co-moving frame of the blob and unprimed quantities refer to the galaxy frame.  If we assume a nominal FSRQ accretion disk luminosity of $L_{disk}\sim10^{45}$ erg s$^{-1}$ and that a fraction $\xi = 0.1$ of this radiation is reprocessed by  both the BLR and the MT, a typical luminosity of the external radiation field is $L_{ext} \sim 10^{44}$ erg s$^{-1}$ \citep{ghisellini09}.

	If the emission site is located within the BLR (at $R \sim 10^{17} \mathrm{cm}$),the photon field can be considered isotropic in the galaxy frame and using Eq. \ref{eq:uiso} its co-moving energy density is 
	
	\begin{equation}
	\label{eq:denseblrb}
	U'_{BLR} \approx \,2.6 \; \Gamma_{10}^{2} L_{BLR, 44} R^{-2}_{BLR,17} \;  \mbox{erg cm}^{-3}, 
	\end{equation}	
where $\Gamma_{10}= 10^{-1} \Gamma,$ $L_{BLR,44}=10^{-44} L_{BLR}$, and $R_{BLR,17}=10^{-17}R_{BLR} .$ Similarly, the  MT photon field is isotropic inside the BLR and its co-moving seed photon energy density is 

	\begin{equation}
	\label{eq:densemt}
	U'_{MT} \approx  2.6 \times 10^{-2} \; \Gamma_{10}^2 L_{MT, 44} R^{-2}_{MT,18} \;  \mbox{erg cm}^{-3},
	\end{equation} 
where $L_{MT,44}=10^{-44}L_{MT}$ and $R_{MT,18}=10^{-18}R_{MT}$. Clearly, inside the $R_{BLR}$ the co-moving BLR photon field energy density  dominates over that of the MT by a factor of $\sim 100$. 
	 
	 If the emission site is located at $R \gtrsim 10^{18} \:\mathrm{cm} $ (within the MT) then the BLR  UV photons enter the emitting region practically from behind, and as per Eq. \ref{eq:ubhd} 
	\begin{equation} 
	\label{eq:denseblrm}
	 U'_{BLR}  \approx 2.6 \times 10^{-4} \; L_{BLR, 44} R^{-2}_{MT,18} \Gamma_{10}^{-2}\; \mbox{erg cm}^{-3}.  
	\end{equation}
	 The IR photons from the MT  retain the same co-moving energy density previously given by Eq. \ref{eq:densemt}.  In this case, therefore, it is the MT  that dominates the co-moving photon energy density.
 
 	These external photon field co-moving energy densities need to be compared to the synchrotron photon field energy density. If $R_{blob}$ is the size of the emitting blob, the co-moving synchrotron photon energy density is $U'_{synch}  \approx  L_{synch}/ ( 4\pi c R_{blob}^2 \Gamma^4)$.  The most plausible assumption for the size of the emitting region, however, is to set an upper limit to it by its variability timescale: $R_{blob}=c t_{var} \delta $.  We then obtain a lower limit for the synchrotron energy density 
	\begin{equation}
	\label{eq:densesynch}
	U'_{synch}  \approx 2.3 \; L_{synch,46} \; t_{var, 1h}^{-2} \; \Gamma_{10}^{-6} \;  \mbox{erg cm}^{-3},
	\end{equation}
where $L_{synch,46}=L_s \times 10^{-46}$, $t_{var,h}=3600 \times t_{var}$ is the observed variation time in hours, and $\Gamma_{10}$ is the same as previously defined. Note that a 6 hour variability scale has been observed by Fermi \citep{abdo10b}.   
		
		Comparison of the external energy density $U_{ext}$ (using Eqs. \ref{eq:uiso} or \ref{eq:ubhd} depending on if the emission is isotropic)  and synchrotron seed photon energy density $U_{synch}$ (Eq. \ref{eq:densesynch}) shows that an isotropic external photon field dominates, i.e $U_{ext}>U_{synch}$, if 
			
	\begin{equation}
	\label{eq:gammalimit}
	\Gamma > 17.5 \left(\frac{L_{synch,46} R_{ext, 18}^2}{t_{var,h}^2 L_{ext,44}}\right)^{1/8}
	\end{equation}  

Using the BLR and MT descriptions adopted above and $t_{var} = 6$ hours, results in a $\Gamma> 6.2$ inside the BLR and $\Gamma > 11.2$ inside the MT, but outside the BLR. 
VLBI studies of superluminal speeds in powerful FSRQs \citep[e.g. see figure 24 of ][]{jorstad05} show that for most FSRQs $10\lesssim \Gamma \lesssim 20$. Therefore, this  condition is  clearly satisfied inside the BLR, and it is expected to be satisfied in the majority of cases in the MT.

 \section{Cooling in the BLR vs Cooling in the MT}
	{\sl The critical difference between the BLR and the MT is the energy of the seed photons: photons originating from the BLR are UV photons ($\epsilon_0 \approx 10^{-5}$) while photons originating from the MT are IR photons ($\epsilon_0 \approx 10^{-7}$), where $\epsilon_0=h \nu/mc^2$.  This difference by a factor of $\sim 10^2$ in typical photon energy is critical in that it affects the energy regime in which electron cooling takes place, and thus the energy dependence of the electron cooling time.}

\begin{figure}
	\includegraphics[width=88mm]{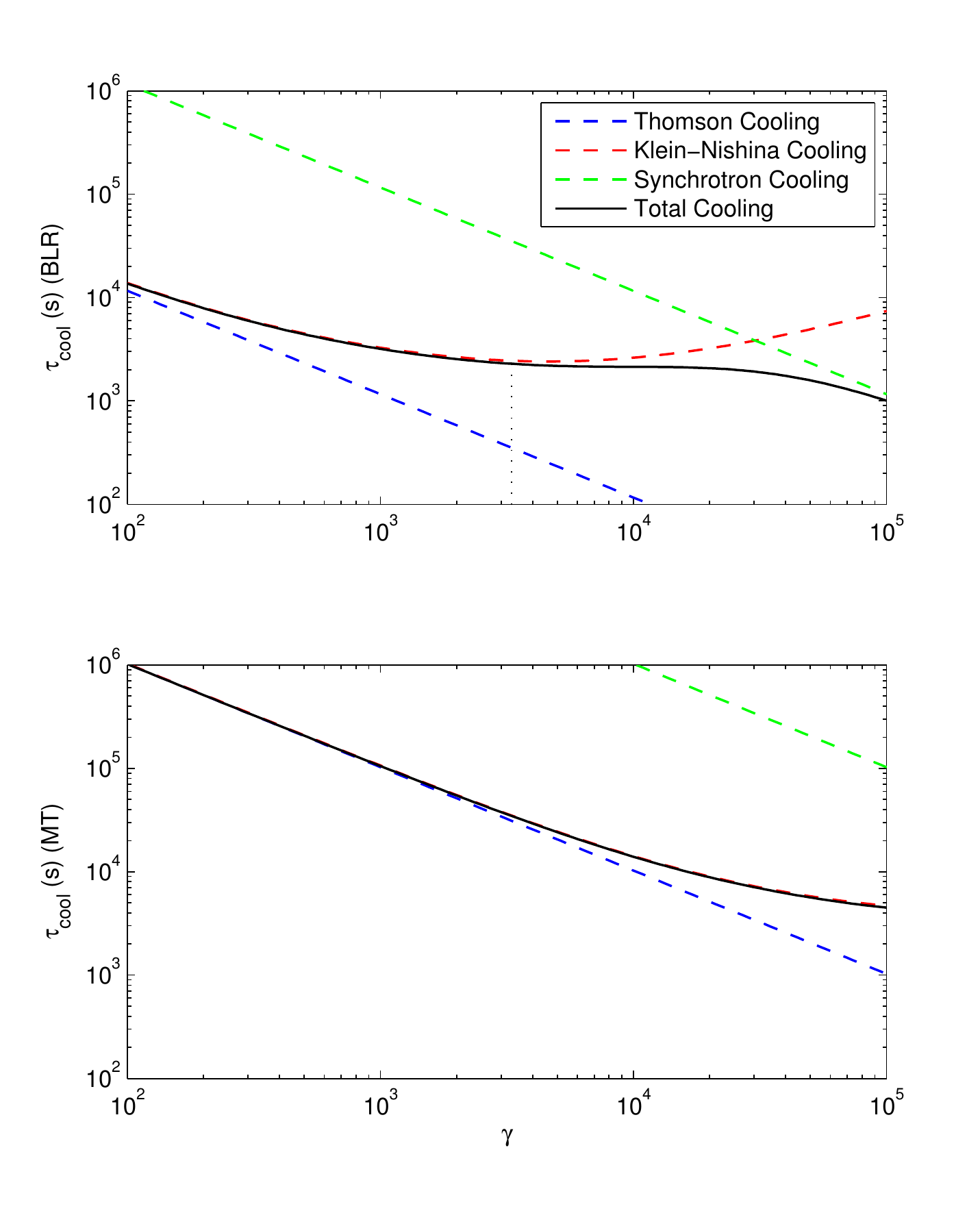}

\caption{Cooling time in the blob frame as a function of co-moving electron energy $\gamma$. Top panel: Blazar emission site located in the BLR.  Bottom panel: Blazar emission site located in the MT.  The dotted lines represent the various cooling mechanisms (blue = Thomson cooling, red = KN cooling, green = synchrotron cooling), the solid black line is the total cooling time. The dotted black line in the top panel indicates where $\gamma \epsilon_{0,BLR} \sim 1.$ Plots were calculated for the following values: seed photon energies: $\epsilon_{0,BLR} = 3 \times 10^{-5}$ and $\epsilon_{0,MT} = 6 \times 10^{-7}$, energy densities $U_{BLR} = 2.65 \times 10^{-2}$ erg cm$^{-3}$, $U_{MT} = 3 \times 10^{-4}$ erg cm$^{-3}$, a $U_B$ that corresponded to a Compton dominance of $\sim 100$, and $\Gamma_{bulk} = 10$. Because BLR photons have energies higher than MT photons by a factor of $\sim 100$, transition from the Thomson to the KN regime occurs at values of $\gamma$ $\sim 100$ times lower in the BLR (top panel)  than in the MT (bottom panel).}

\label{fig:tblrmt}
\end{figure}

In powerful FSRQs the IC emission in high states can dominate over the synchrotron emission by a factor of up to $\sim100$ \citep[][]{abdo10a}. In cases of high Compton dominance, the primary electron cooling mechanism is IC scattering.  For electrons cooling in the Thomson regime ($\gamma\epsilon_0 \lesssim 1$, where both the electron Lorentz factor $\gamma$ and the seed photon energy $\epsilon_0$ are measured in the same frame) the cooling rate $\dot{\gamma} \propto \gamma^2$.  For electrons with $\gamma \epsilon_0 \gtrsim 1$ cooling takes place in the Klein-Nishina (KN) regime with $\dot{\gamma} \propto \ln \gamma$ \citep{blum70}.  The electron cooling time given by $\gamma/\dot{\gamma}$ is then $\propto \gamma^{-1}$ in the Thomson regime and $\propto \gamma/\ln\gamma$ in the deep KN regime. In the transition from Thomson to KN regimes (around $\gamma \epsilon_0 \sim 1$) the cooling time flattens and becomes essentially energy independent (Fig \ref{fig:tblrmt}). 

	The effects of the transition between Thomson and KN regimes on the electron energy distribution (EED) and the resultant spectrum of the synchrotron and IC emission have been studied before \citep[e.g.][]{blum71,zdz89,dermer93,sok04,mod05,kusun05,georganopoulos06,manolakou07,sikora09}.  In short, because $\dot{\gamma} \propto \gamma^2$ in the Thomson regime and $\dot{\gamma} \propto \ln\gamma$ in the KN regime, the cooling time $\tau_{cool}' = \gamma/\dot{\gamma}$ scales as $\gamma^{-1}$ in the Thomson regime and as $\gamma/\ln\gamma$ in the KN regime.  

	 Because  the cooling time  is approximately energy independent around $\gamma \epsilon_0 \sim 1$, this energy-independent cooling time will be manifested at energies lower by a factor of  $\sim 100$  for cooling taking place in the BLR compared to cooling taking place  in the MT, since the BLR seed photons have an energy higher than that of the MT by a factor of $\sim 100$. This can be seen in Fig. \ref{fig:tblrmt} where we plot the electron cooling time for a source with a ratio of external photon field energy density  $U_0'$ in the co-moving frame to co-moving magnetic field energy density $U_B$,  $U_0'/U_B=100$. This corresponds to a factor of $\sim 100$ Compton dominance (ratio of inverse Compton to Synchrotron luminosity), similar to what is observed in the most Compton dominated sources.

	The transition from Thomson cooling to KN cooling, and from KN cooling to synchrotron cooling can also be seen in Fig. \ref{fig:tblrmt}.  If the electrons are cooling on photons from the BLR, cooling takes place in the Klein-Nishina regime and the cooling time scale is approximately energy independent around $\gamma \epsilon_0 \sim 1$ (Fig. \ref{fig:tblrmt}, top panel).  If the electron population cools on photons from the molecular torus, cooling takes place in the Thomson regime (Fig. \ref{fig:tblrmt}, bottom panel).  The cooling time is heavily energy dependent, and any variations should consequently exhibit heavy energy-dependence.  

\section{The Diagnostic Test}
	The energy dependence of the cooling time results in an energy dependence (or lack thereof) of variations: if the blazar emission site is located within the BLR, variations should be achromatic.  The energy dependence of the variations can be used as a diagnostic test to determine if the GeV emission site is located within the BLR.  By comparing \textit{Fermi} light curves of flares at different energies, we propose that the energy dependence of the light curve can be used as a diagnostic test to rule out whether the GeV flare originates in the BLR or MT.

\subsection{Numerical Simulation Results}
	To demonstrate the effect of the energy independence or dependence of the electron cooling time on the variability of a flare, we utilized a one-zone numerical model to simulate a flare. We initialized the code with values appropriate for a high power blazar with a Compton dominance of $\sim 100$.  For this particular simulation we assumed a source size $R=10^{16} \mathrm{cm}$, bulk Lorentz factor $\Gamma = 10$, co-moving injected electron luminosity $L = 2 \times 10^{44}$ erg s $^{-1}$, maximum electron Lorentz factor $\gamma_{max} = 10^5$, and electron index $p=2.5$.  For the case of a flaring region located within the BLR we assumed an initial photon energy $\epsilon_0 = 3 \times 10^{-5}$ and an energy density (in the galaxy frame) $U_{BLR}=2.6 \times 10^{-2}$ erg cm$^{-3}$.  For the case of a flaring region located outside the BLR we assumed an initial photon energy $\epsilon_0 = 6 \times 10^{-7}$ and an energy density $U_{MT}=3 \times 10^{-4}$ erg cm$^{-3}$.  For each case the magnetic field \textbf{B} was fixed to assume a Compton dominance $U_{EC}/U_B=100$.
	
\begin{figure}
	\includegraphics[width=87mm]{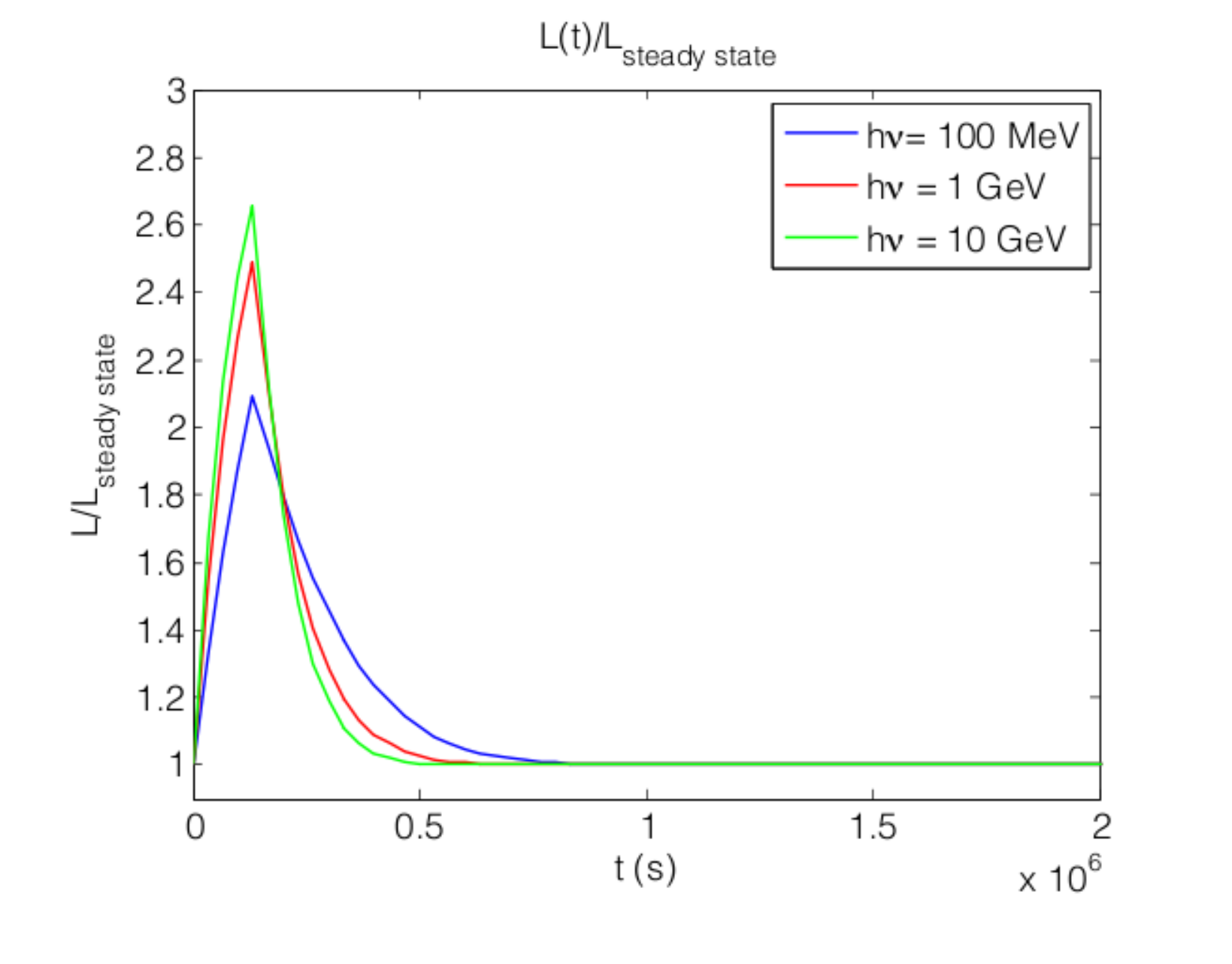}
	\caption{Light curve at various energies (BLR seed photons): $\epsilon_0=100$ MeV, $\epsilon_0=1$ GeV , 
	$\epsilon_0=10$ GeV. }
	\label{fig:blrflare}
\end{figure}

\begin{figure}
	\includegraphics[width=87mm]{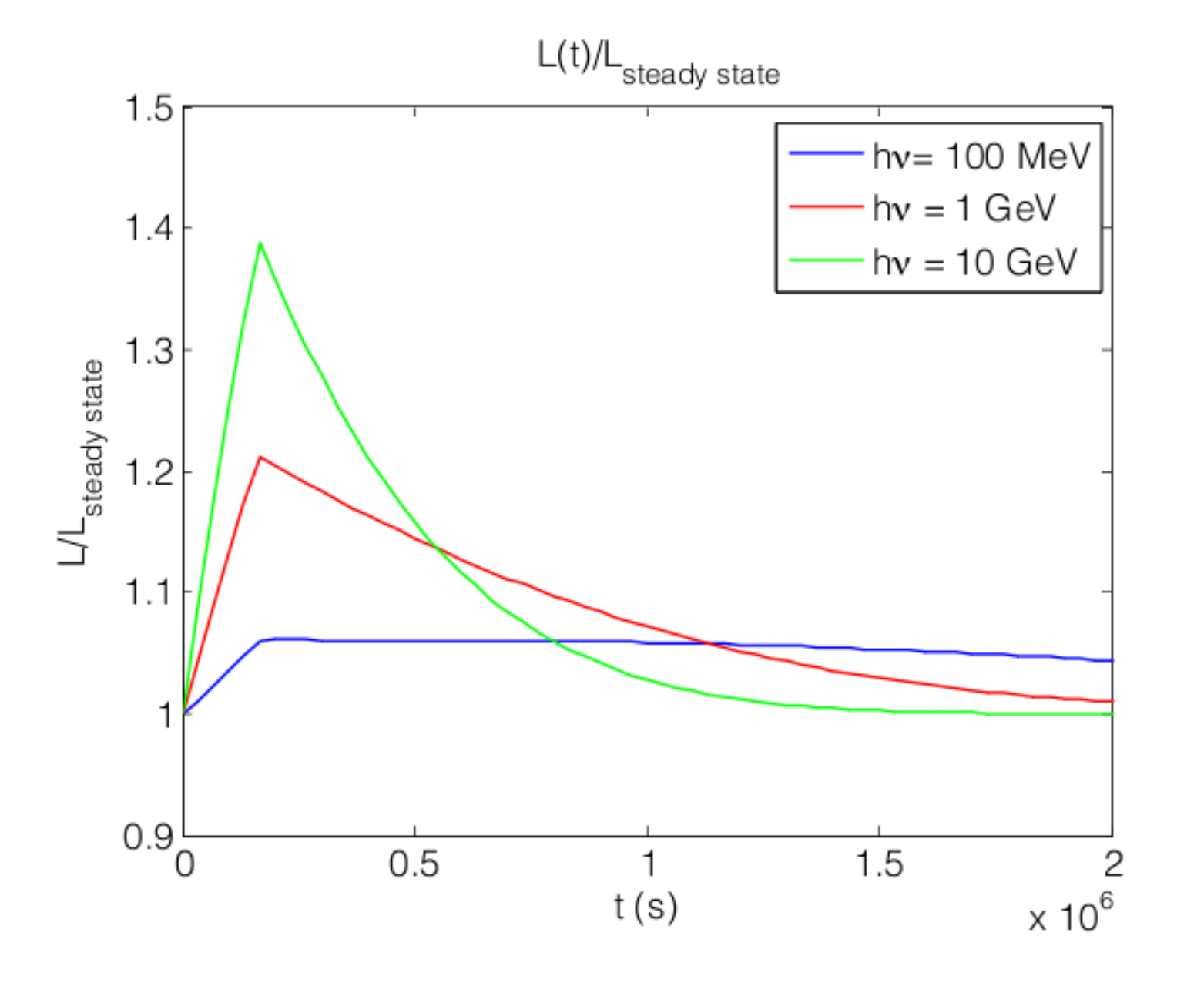}
	\caption{Light curve at various energies (MT seed photons): $\epsilon_0=100$ MeV, $\epsilon_0=1$ GeV, $\epsilon_0=10$ GeV. }
	\label{fig:mflare}
\end{figure}

	The system behaves as expected, showing a noticeable difference in the decay rate as well as the amplitude of the flare depending on if the seed photons originated from within or from outside the BLR (see Figs. \ref{fig:blrflare} and \ref{fig:mflare}).  This difference in decay rate and amplitude can be used as a diagnostic test to differentiate between flares that take place inside or outside the BLR.  

\subsection{Light-Crossing Time Effects}
	Because the GeV emitting region is not a point source, any change in the luminosity of the blob will not be seen instantaneously.  Instead, the observed decay time of the light curve is the result of the actual decay time and the light-crossing time inherent in the blob.  To test whether the light-crossing time would erase any difference in the light curves (as predicted by our diagnostic) for the case of the emitting region being located outside the BLR, we modeled the flux of a source with light-crossing time $t_{LC}$ that is decreasing in flux.  We assumed the flux of the source is decaying exponentially, $F(t) = F_0 \exp^{-t/t_c}$, where $t_c$ is the cooling time at a specific energy.  
	
	Our diagnostic predicts differences in the decay time of the light curves for the case where the emitting region is located within the MT; for the purposes of this demonstration we assume cooling occurs in the Thomson regime and as a result $t_c \propto \epsilon^{-1/2}$.  We plot the resultant light curves for cooling times differing by a factor of $1/\sqrt{10}$ (i.e. energies differing by a factor of 10).
\begin{figure}
	\includegraphics[width=85mm]{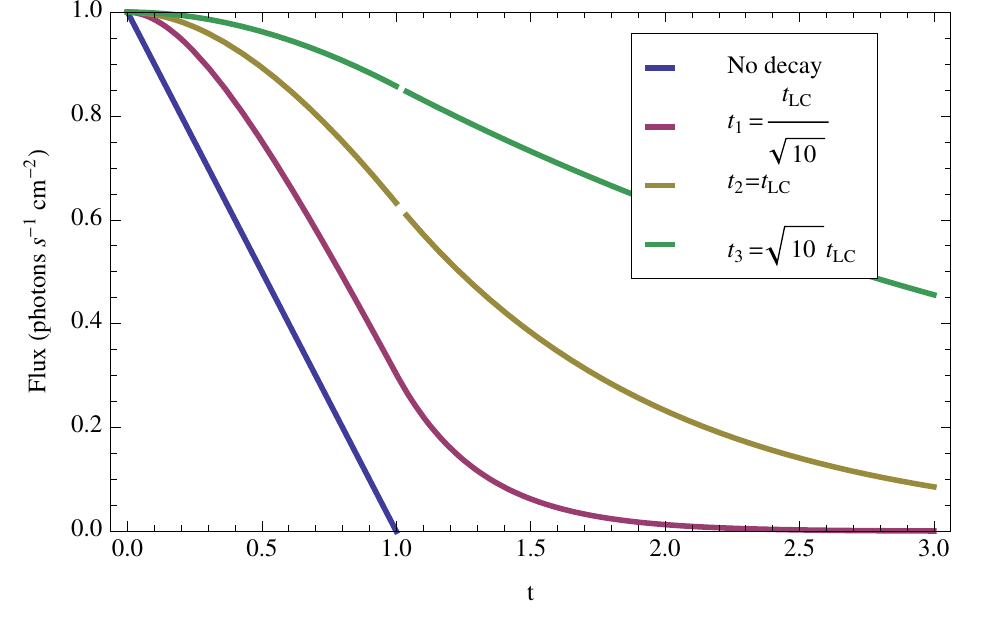}
	\caption{Light curves with light-crossing time effects factored in.  Initial flux and light-crossing time are normalized to 1. Light curves are plotted for energies differing by a factor of 10.}
	\label{fig:lctime}
\end{figure}
As evident in Fig. \ref{fig:lctime}, even with light-travel time effects convolved with exponential decay times, the differences in the decay times are still manifested.

\subsection{Cooling time difference constraints on  {\boldmath $U_{MT}$} and  {\boldmath $\Gamma$}.}
\begin{figure}
\includegraphics[width=90mm]{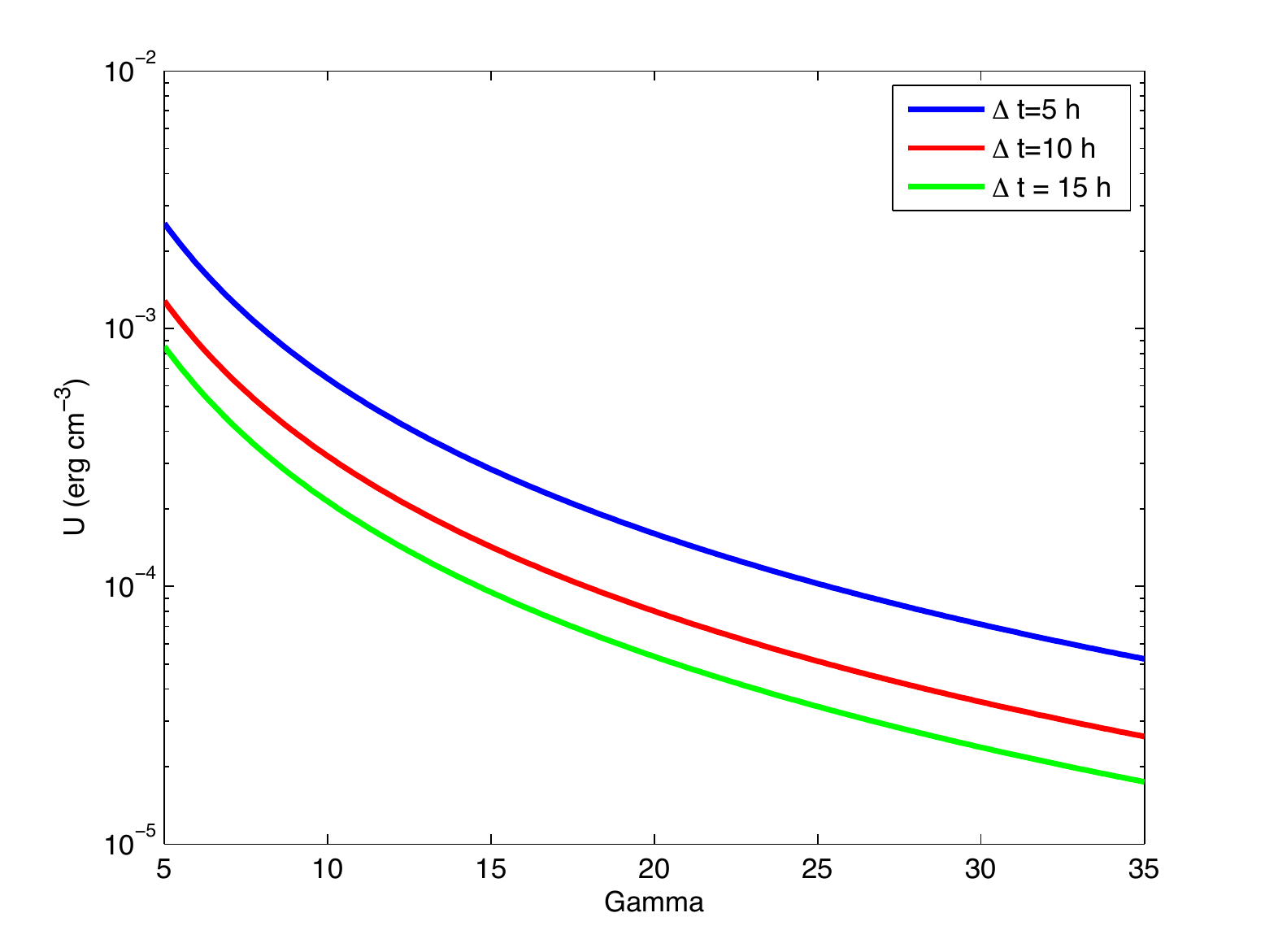}
\caption{Upper limit in the MT energy density ($U$) as a function of bulk Lorentz factor $\Gamma$ plotted for $\Delta t$ corresponding to 5, 10, and 15 h.  Accepted values of of $U$ and $\Gamma$ values fall below the curves.}
	\label{fig:ugam}
\end{figure}

		A flare that occurs in  the MT should exhibit energy dependent cooling  times because cooling occurs in the Thomson regime.  Practical application of the diagnostic hinges on the requirement that the decay times in different energy bands should have  detectable differences.  In the Thomon regime, the cooling time is given by $t_c = frac{3 m_e c }{4 U \Gamma^2 \sigma_T}\frac{\epsilon_0}{\epsilon}^{1/2}$, where $\epsilon_0$ is the seed photon energy and $\epsilon$ is the photon energy after IC scattering. The expected difference in cooling time at two different energies is then
		\begin{equation}
		\Delta t = \frac{3 m_e c \epsilon_0^{1/2}}{4 U \Gamma^2 \sigma_T}\left(\epsilon_{LE}^{-1/2} - \epsilon_{HE}^{-1/2}\right),
		\end{equation}
		where $\epsilon_{LE}$ and $\epsilon_{HE}$ are the photon energies at which we calculate the cooling time, and $U$ is the MT energy density.  
Because of {\sl Fermi's} three hour time resolution, $\Delta t\gtrsim 3$ h
	and this sets an upper limit to the product $U\Gamma^2$.	
		  The $\Gamma$-$U$ parameter space    for which the difference in decay times are distinguishable is plotted in Fig. \ref{fig:ugam} for three plausible values of $\Delta t$, with the area below each line corresponding to the acceptable part of the parameter space. The permitted part of this diagram comfortably accommodates
		  bulk Lorentz factors between 10 and 20 and MT photon densities $\sim 10^{-4}$ erg cm$^{-3}$.

\subsection{SSC in the MT?}

Our discussion following equation (\ref{eq:gammalimit}) suggests that it is possible that in the MT, for relatively low values of $\Gamma$, $U_{synch} > U_{EC}$ and the dominant photon field is from jet synchrotron photons. Even if this is the case, cooling still occurs in the Thomson regime and the energy dependent cooling time is still manifested. For FSRQs,  the peak of the synchrotron spectral energy distribution is at $\nu_s \sim 10^{13}$ Hz \cite{giommi11}, which in the comoving frame corresponds to $\epsilon_s =h \nu_s/ (m_ec^2\Gamma)\approx 10^{-7}/\Gamma$. These seed photons, therefore, have a lower energy than the energy of the MT photons in the comoving frame, and will  be deeper in the Thomson regime compared to the MT photons. Because cooling still occurs in the Thomson regime, the energy dependent cooling time, therefore, is expected to be present, even in the case that the dominant photon density is that of the synchrotron photons.

\section{Conclusions}
	We have presented a diagnostic test that utilizes blazar variability to determine the location of the GeV emitting site in blazars.  The energy difference in seed photons originating from the BLR versus seed photons originating from the MT causes electrons within the emitting site to cool in different energy regimes.  
	
	For the case where the GeV emitting site is located within the BLR, cooling takes place at the onset of the KN regime, and the resultant electron cooling time is energy-independent.  We have demonstrated that the associated light curves exhibits decay times that are approximately energy independent.  Conversely, for the case where the GeV emitting site is located outside the BLR, cooling takes place in the Thomson regime and the electron cooling times are heavily energy dependent.  In this case, the associated light curves exhibit energy dependence of their decay times.  
	
	The energy dependence of the decay time of the light curves is visible within the \textit{Fermi} energies; these differences can be used as a diagnostic test to determine whether the GeV emitting region is located inside or outside the BLR.  These effects are observable within the time resolution of $\sim$ few hours that  \textit{Fermi} has achieved for bright flares and are not erased due to light-travel time effects.   Light curves from a sufficiently bright and rapid flare \cite[such as that in 3C 454.3;][]{abdo3c454} should be compared at different energies.   If the GeV emitting site is located within the BLR,  the decay times will exhibit no energy dependence, whereas if the emitting site is located within the MT, the decay times will exhibit energy dependence.

\end{document}